\documentclass[11pt,twocolumn]{article}
\usepackage[a4paper, margin=0.75in]{geometry}
\usepackage{amsmath, amssymb, amsthm}
\usepackage{graphicx}

\usepackage{dcolumn}
\usepackage{bm}
\usepackage[utf8]{inputenc}
\usepackage[T1]{fontenc}
\usepackage{mathptmx}
\usepackage{etoolbox}
\usepackage {cite}
\usepackage{enumitem}

\title{Programmable time-frequency mode encoded quantum state generator for silicon-on-insulator platform}
\author{Liao Ye,$^{1}$ Haoran Ma,$^{1}$ Fanjie Ruan,$^{1}$ Yuehai Wang,$^{1}$ and Jianyi Yang$^{1,*}$ 
	\\Institute of Microelectronic and Nanoelectronics, 
	\\College of Information Science and Electronics Engineering, 
	\\Zhejiang University, Hangzhou 310027, China
	\\12031018@zju.edu.cn}
\date{\today}

\begin{document}
	\begin{sloppypar}
		\maketitle
\begin{abstract}
	We propose a method for the programmable generation of time-frequency mode (TFM) encoded quantum states of light on the silicon-on-insulator (SOI) platform. The state generator consists of an N-tap finite impulse response filter and a Mach-Zehnder interferometer (MZI)-based coupled-ring resonator. Through numerical simulations, its capability of producing TFM-encoded maximally entangled states in two, three, and four dimensions is theoretically demonstrated, with fidelities of 0.950, 0.954, and 0.971, respectively.
\end{abstract}

\section{Introduction}
\noindent High-dimensional entangled states of light have recently gained attention for their impact on the advancement of quantum computation\cite{chi2022programmable} and communication protocols\cite{cerf2002security}. Photons are excellent carriers of quantum information due to their encodable degrees of freedom, such as path\cite{chi2022programmable,zheng2023multichip}, time-frequency\cite{imany201850,clementi2023programmable}, and transverse spatial modes\cite{feng2022transverse}. Among these, time-frequency degree of freedom is distinguished by its inherent high-dimensional nature, compatibility with spatially single-mode devices\cite{ansari2018tailoring}, and insensitivity to polarization-mode dispersion during transmission \cite{zhang2008distribution}. Time-frequency mode (TFM) encoding\cite{brecht2015photon}, which encodes quantum information in the spectral envelope of single-photon wave packets\cite{graffitti2020direct}, enables an infinite number of encodable dimensions within specific temporal or spectral ranges\cite{ansari2018tailoring,law2000continuous}.

Controlled generation of TFM-encoded states has been demonstrated on platforms with second-order optical nonlinearity, specifically in the production of pure states\cite{ansari2018heralded} and the biphoton Bell states $\left| {{\psi }^{+}} \right\rangle $ \cite{ansari2020remotely} and $\left| {{\psi }^{-}} \right\rangle $\cite{graffitti2020direct,francesconi2020engineering}. These generators rely on a spatial light modulator (SLM) for spectral\cite{ansari2018heralded,ansari2020remotely} or spatial\cite{francesconi2020engineering} waveform shaping of the pump, or on domain-engineered crystals to tailor the phase-matching function\cite{graffitti2020direct}, limiting their integrability or programmability. Despite the advanced programmability of integrated photonic platforms such as SOI platform, for the time-frequency degree of freedom of light, previous work has mainly focused on producing heralded single photons with high spectral purity\cite{paesani2020near,burridge2023integrate} or discrete frequency-bin encoded quantum states\cite{imany201850,clementi2023programmable}. To the best of our knowledge, the controlled generation of TFM-encoded entangled states on the SOI platform has not yet been explored.

In this letter, we present a method for the programmable generation of TFM-encoded quantum states of light on the SOI platform without post-manipulation. The proposed generator consists of an N-tap finite impulse response (FIR) filter and a coupled-ring resonator, which respectively shape the spectral waveform of the pump pulses and implement reconfigurable field enhancement functions. Inversion procedures for model parameter retrieval are employed to optimize the fidelity of the generated states. Numerical simulations demonstrate the generation of a TFM-encoded Bell state $\left| {{\phi }^{-}} \right\rangle $ with a fidelity of 0.950, as well as three- and four-dimensional maximally entangled states with fidelities of 0.954 and 0.971, respectively.

\section{Device configuration}
\begin{figure}[t]
	\centering
	\includegraphics[width=\linewidth]{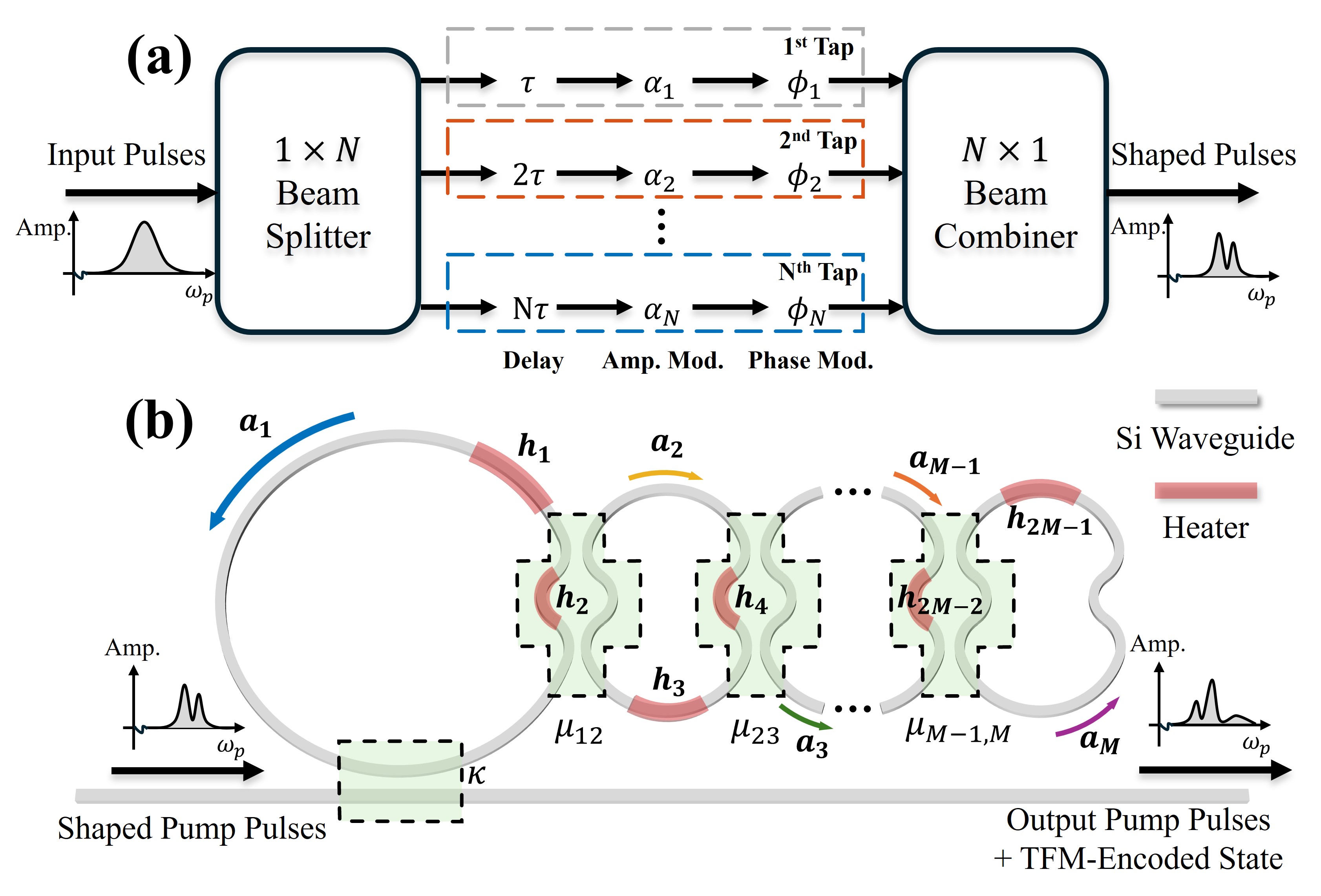}
	\caption{Schematic configuration of (a) the N-tap FIR filter and (b) the coupled-ring resonator-based photon-pair source.}
	\label{fig:1}
\end{figure}

The schematic configuration of the N-tap FIR structure is shown in Fig.~\ref{fig:1} (a), which has previously been demonstrated as an arbitrary waveform generator in Ref.\cite{liao2015arbitrary}. Here, we expand the number of taps to $N$ to enhance the spectral complexity of the shaped pulses. The input pulse is equally divided into $N$ paths using a $1\times N$ beam splitter. These paths introduce delays from $\tau$ to $N\tau$, followed by independent amplitude and phase modulation for each tap, where $\tau$ represents the base delay time. These modulated pulses are coherently combined using an $N\times 1$ beam combiner. Applying Mason's gain formula\cite{mason1953feedback}, the frequency-domain transfer function of the FIR pulse shaper can be expressed as\cite{liao2015arbitrary}
\begin{equation}
	\label{Eq:1}
	H\left( \omega  \right)=\sum\limits_{n=1}^{N}{{{\alpha }_{n}}\exp \left[ j\left( {{\phi }_{n}}-n\omega \tau  \right) \right]},
\end{equation}
where ${\alpha }_{n}$ and ${\phi }_{n}$ are the amplitude and phase modulation factors of the n-th tap, respectively. Tuning the parameters ${\alpha}_{n}$ and ${\phi}_{n}$ allows for spectral shaping of the pump pulses.

\begin{figure}[t]
	\centering
	\includegraphics[width=\linewidth]{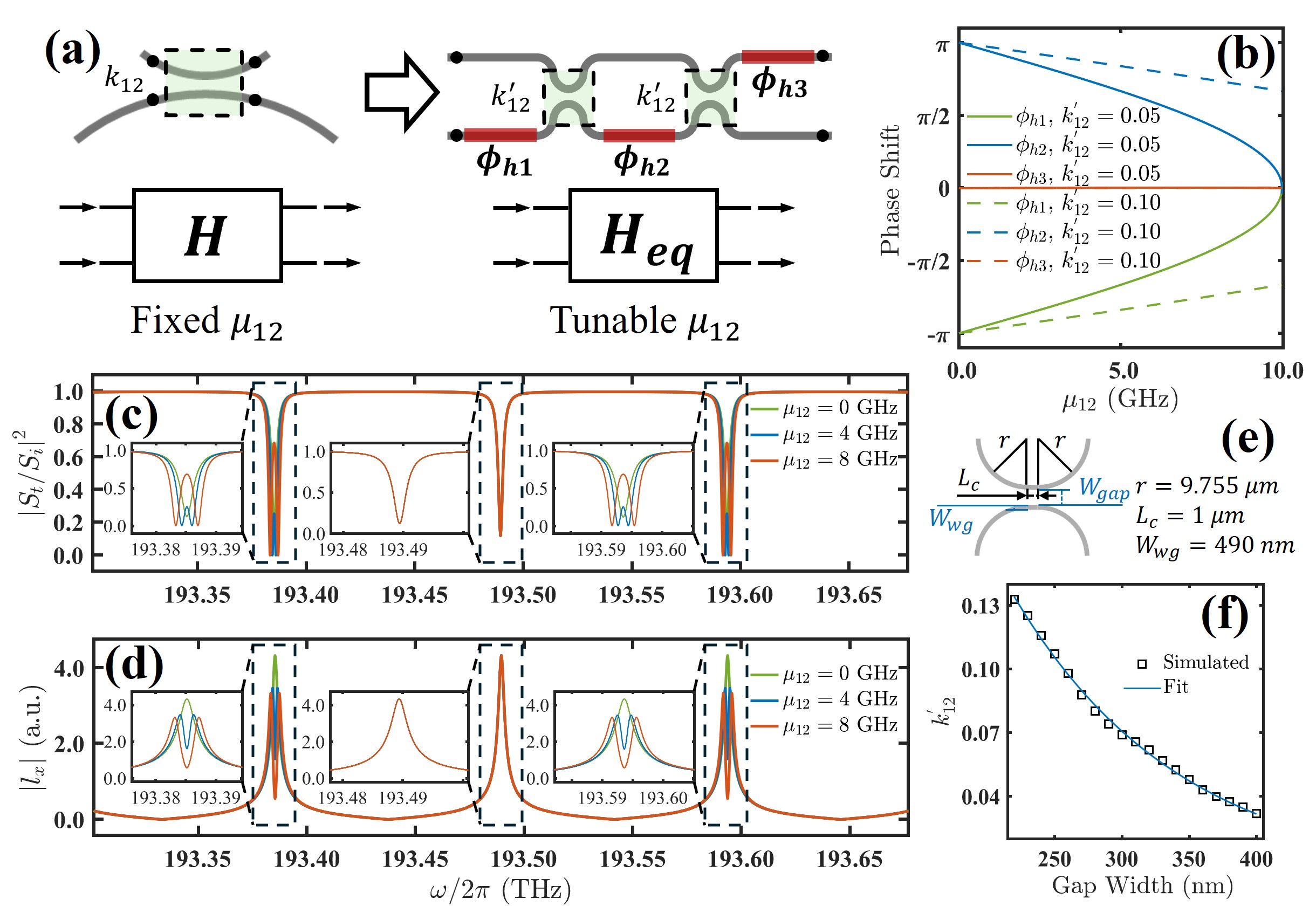}
	\caption{(a) Schematic of the MZI coupler with a tunable ${{\mu}_{12}}$. (b) Phases shifts ${{\phi }_{h1}}$, ${{\phi }_{h2}}$, and ${{\phi }_{h3}}$ as functions of ${{\mu}_{12}}$. (c) and (d) Spectra of (c) ${{\left| {{S}_{t}}/{{S}_{i}} \right|}^{2}}$ and (d) ${{\left| {l}_{x} \right|}}$ at ${\mu}_{12}$ values of 0, 4, and 8 GHz. Insets provide magnified views of the spectra at the resonant frequencies. (e) Geometry of the coupling region in the MZI coupler, where $r$ is the bend radius, $L_c$ is the coupler length, and $W_{wg}$ is the waveguide width. (f) Simulated and fitted $k_{12}'$ as functions of the gap width.}
	\label{fig:2}
\end{figure}

The spectrally shaped pump is then directed to a coupled-ring resonator, which allows the nonlinear process to be configurable. Figure~\ref{fig:1} (b) shows the schematic of the photon-pair source, consisting of M-stage coupled rings and a single-mode bus waveguide. The first-stage larger main ring is coupled to the bus, and the smaller auxiliary rings in the remaining $\text{M-1}$ stages are coupled to the previous stage via tunable Mach-Zehnder interferometer (MZI) couplers. The free spectral range (FSR) ratio of the main and auxiliary rings is set to $1:2$, resulting in splitting at resonances separated by two FSRs of the main ring. For a two-stage coupled-ring resonator with a degenerate resonant frequency ${\omega}_{0}$ and an input optical field ${{S}_{i}}$, the time-domain behavior can be described as follows\cite{little1997microring,fan2003temporal}:
\begin{eqnarray}
	\label{Eq:2}
	\frac{d{{a}_{1}}}{dt}=i{{\omega }_{0}}{{a}_{1}}-\frac{1}{{{\tau }_{1}}}{{a}_{1}}-i{{\mu}_{12}} {{a}_{2}}-i\kappa {{S}_{i}},~~~~~~~\nonumber\\
	\frac{d{{a}_{2}}}{dt}=i{{\omega }_{0}}{{a}_{2}}-\frac{1}{{{\tau }_{2}}}{{a}_{2}}-i{{\mu}_{12}} {{a}_{1}},~~~~{{S}_{t}}={{S}_{i}}-i\kappa a,
\end{eqnarray}
where ${a}_{1}$ and ${a}_{2}$ are the complex amplitudes normalized to the optical energy stored in the main and auxiliary rings, respectively, and they are related to the intra-cavity power flow amplitudes ${A}_{1}$ and ${A}_{2}$ through $\left| {{a}_{m}} \right|=\left| {{A}_{m}} \right|\sqrt{{L}_{m}/{{v}_{g}}}$ (where ${m} = {1,2}$)\cite{little1997microring}. Here, ${L}_{m}$ is the perimeter of the ring, and ${v}_{g}$ is the group velocity. The parameters $1/{{\tau}_{1}}$ and $1/{{\tau}_{2}}$ represent the amplitude decay rates associated with ${a}_{1}$ and ${a}_{2}$, respectively. ${\kappa}$ is the field coupling coefficient between the bus and the main ring, and ${{S}_{t}}$ represents the transmitted wave at the bus's output port. ${{\mu}_{12}}$ is the mutual energy coupling coefficient between modes ${a}_{1}$ and ${a}_{2}$, which is related to the power coupling coefficient ${{k}_{12}}$ through ${{\mu }_{12}}={{k}_{12}}\sqrt{v_{g}^{2}/\left( {{L}_{1}}{{L}_{2}} \right)}$\cite{little1997microring}. By defining ${{l}_{x}}\left( \omega  \right)={{A}_{1,x}}\left( \omega  \right)/{{S}_{i}}\left( \omega  \right)$ (where ${x} = {i,p,s}$), the field enhancement for the split resonance $x$ can be expressed as
\begin{eqnarray}
	\label{Eq:3}
	{{l}_{x}}\left( \omega
	\right)=\sqrt{\frac{{{v}_{g}}}{{{L}_{1}}}}\frac{{{a}_{1,x}}(\omega )}{{{S}_{i}}(\omega )}~~~~~~~~~~~~~~~~~~~~~~~~~~~~~~~~~~~~~~~~~~~~~~~~~~~~~~~~~~~~~\nonumber\\
	=\sqrt{\frac{{{v}_{g}}}{{{L}_{1}}}}\frac{{{\kappa }_{x}}\left[ \left( \omega -{{\omega }_{x}} \right)-i/{{\tau }_{2,x}} \right]}{\left[ i\left( \omega -{{\omega }_{x}} \right)+1/{{\tau }_{1,x}} \right]\left[ i\left( \omega -{{\omega }_{x}} \right)+1/{{\tau }_{2,x}} \right]+\mu _{12,x}^{2}}.
\end{eqnarray}
For degenerate resonant frequencies of the main and auxiliary rings, increasing ${\mu}_{12}$ causes significant spectral splitting in ${{\left| {{S}_{t}}/{{S}_{i}} \right|}^{2}}$ and ${{\left| {l}_{x} \right|}}$, as shown in Figs.~\ref{fig:2}(c) and ~\ref{fig:2}(d). In contrast, for non-degenerate resonant frequencies, the ${\mu}_{12}$ in Eq.~(\ref{Eq:3}) is set to zero, and ${{\left| {l}_{x} \right|}}$ exhibits a Lorentzian line shape\cite{fan2003temporal}. Given that the FSR of the auxiliary ring is twice that of the main ring, selecting three adjacent resonances of the main ring for the idler, pump, and signal allows for the generation of entangled states in subsequent simulations. The reconfiguration of ${{\mu}_{12}}$ is provided by the tunable MZI coupler, which consists of two identical coupling regions with a power coupling coefficient $k_{12}'$ and three phase shifts ${{\phi }_{h1}}$, ${{\phi }_{h2}}$, and ${{\phi }_{h3}}$. This coupler functions as an equivalent point coupler with a tunable ${\mu}_{12}$\cite{zhao2024clements}, as sketched in Fig.~\ref{fig:2}(a). Additionally, Fig.~\ref{fig:2}(b) demonstrates the phase adjustment finesse $\left| d\phi /d{{\mu }_{12}}\right|$ of the MZI coupler within a tuning range of $\mu_{12}$ from 0 to 10 GHz. When $k_{12}'={0.05}$, the finesse is higher than when $k_{12}'={0.10}$, with ${\mu}_{12}$ increasing by approximately 0.08 GHz for each one-degree phase change. Figure.~\ref{fig:2}(e) shows a feasible design for the coupling region geometry. Simulations using Ansys Lumerical FDTD demonstrate that $k_{12}'$ below 0.05 can be achieved when $W_{gap}$ exceeds 345 nm, as shown in Fig.~\ref{fig:2}(f).

\section{Principles}
Disregarding multipair generation and using a micro-ring resonator at the generation stage, the broadband joint spectral amplitude (JSA) function $F({{\omega }_{s}},{{\omega }_{i}})$ of a spontaneous four-wave mixing (SFWM) process, which represents the probability amplitude distribution of signal-idler photon pairs in $\left\{ {{\omega }_{s}},{{\omega }_{i}} \right\}$ space, is constrained to be a narrow-band function\cite{law2000continuous,vernon2017truly}:
\begin{eqnarray}
	\label{Eq:4}
	F({{\omega }_{s}},{{\omega }_{i}})=\int{d{{\omega }_{p}}{{\alpha }_{p}}\left( {{\omega }_{p}} \right)}{{l}_{p}}\left( {{\omega }_{p}} \right){{\alpha }_{p}}\left( {{\omega }_{s}}+{{\omega }_{i}}-{{\omega }_{p}} \right)\nonumber\\
	\times{{l}_{p}}\left( {{\omega }_{s}}+{{\omega }_{i}}-{{\omega }_{p}} \right){{\phi}_{PM}}\left({{\omega }_{p}},{{\omega }_{s}},{{\omega }_{i}}\right){{l}_{s}}\left( {{\omega }_{s}} \right){{l}_{i}}\left( {{\omega }_{i}} \right),
\end{eqnarray}
Here, ${{\alpha }_{p}}\left( {{\omega}} \right)$ is the spectral envelope of the pump, given by ${{\alpha }_{p}}\left( \omega  \right)={{\alpha }_{0}}\left( \omega  \right)H\left( \omega  \right)$, where ${{\alpha }_{0}}\left( {{\omega}} \right)$ is the spectral envelope of the input Gaussian pulse with a full width at $\sqrt{1/e}$ maximum $2{\sigma}_p$, and $H\left( \omega  \right)$ is the transfer function defined in Eq.~(\ref{Eq:1}). ${{\phi}_{PM}}\left({{\omega }_{p}},{{\omega }_{s}},{{\omega }_{i}}\right)$ is the phase-matching function (PMF) determined by the wavevectors of four interacting optical fields (see Eq. ~\ref{Eq:S1} and \ref{Eq:S2} in Appendix A).

Given the achieved programmability of both ${{l}_{x}}\left( \omega  \right)$ and ${{\alpha }_{p}}\left( {{\omega}} \right)$, the JSA in Eq.~(\ref{Eq:4}) can be represented as a product of three intermediary functions: the phase-matching function ${{\phi}_{PM}}\left({{\omega }_{p}},{{\omega }_{s}},{{\omega }_{i}}\right)$, the anti-diagonal pump function $ADP\left( {{\omega }_{p}} \right)$, and the two-dimensional signal-idler function $TDSI\left( {{\omega }_{s}},{{\omega }_{i}} \right)$. The cross-section of an SOI strip waveguide is optimized to $\text{490 nm}\times \text{220 nm}$ (see Appendix A), achieving a phase-matching function of the form ${{\phi }_{PM}}\left( {{\omega }_{s}}-{{\omega }_{i}} \right)$, which provides flexibility for generating the TFM-encoded entangled states\cite{ansari2018tailoring}. The ADP function, which characterizes the anti-diagonal distribution of JSA, is defined as
\begin{equation}
	\label{Eq:5}
	ADP\left( {{\omega }_{p}} \right)=\left[ {{\alpha }_{p}}\left( {\omega} _{p}  \right){{l}_{p}}\left( {{\omega }_{p}} \right) \right]*\left[ {{\alpha }_{p}}\left( {\omega} _{p}  \right){{l}_{p}}\left( {{\omega }_{p}} \right) \right],
\end{equation}
while the $TDSI\left( {{\omega }_{s}},{{\omega }_{i}} \right)={{l}_{s}}\left( {{\omega }_{s}} \right){{l}_{i}}\left( {{\omega }_{i}} \right)$ serves as a two-dimensional filtering function in $\left\{ {{\omega }_{s}},{{\omega }_{i}} \right\}$ space.

To demonstrate the generator's capability and programmability in producing TFM-encoded quantum states, we simulate the generation of TFM-encoded maximally entangled states in two, three, and four dimensions, defined as
\begin{eqnarray}
	\label{Eq:6}
	\left| {{\phi }^{- }} \right\rangle =\frac{1}{\sqrt{2}}\left( {{\left| 0 \right\rangle }_{s}}{{\left| 0 \right\rangle }_{i}}- {{\left| 1 \right\rangle }_{s}}{{\left| 1 \right\rangle }_{i}} \right),~~~~~~~~~~~~~~~~\nonumber\\
	\left| {{\Psi }_{D=3}}  \right\rangle =\frac{1}{\sqrt{3}}\left( {{\left| 0 \right\rangle }_{s}}{{\left| 0 \right\rangle }_{i}}-{{\left| 1 \right\rangle }_{s}}{{\left| 1 \right\rangle }_{i}}+{{\left| 2 \right\rangle }_{s}}{{\left| 2 \right\rangle }_{i}} \right),~~~~~~\nonumber\\
	\left| {{\Psi }_{D=4}}  \right\rangle =\frac{1}{2}\left( {{\left| 0 \right\rangle }_{s}}{{\left| 0 \right\rangle }_{i}}-{{\left| 1 \right\rangle }_{s}}{{\left| 1 \right\rangle }_{i}}+{{\left| 2 \right\rangle }_{s}}{{\left| 2 \right\rangle }_{i}}-{{\left| 3 \right\rangle }_{s}}{{\left| 3 \right\rangle }_{i}}  \right).
\end{eqnarray}
Here, ${{\left| k \right\rangle }_{x}}$ represents the ${x}$ photon (${x} = {s,i}$) occupied by the k-th order (${k} = {0,1,2,3}$) Hermite-Gaussian (HG) modes (see Eq. ~\ref{Eq:S3} in Appendix B) in the frequency domain. 

\begin{figure}[t]
	\centering
	\includegraphics[width=\linewidth]{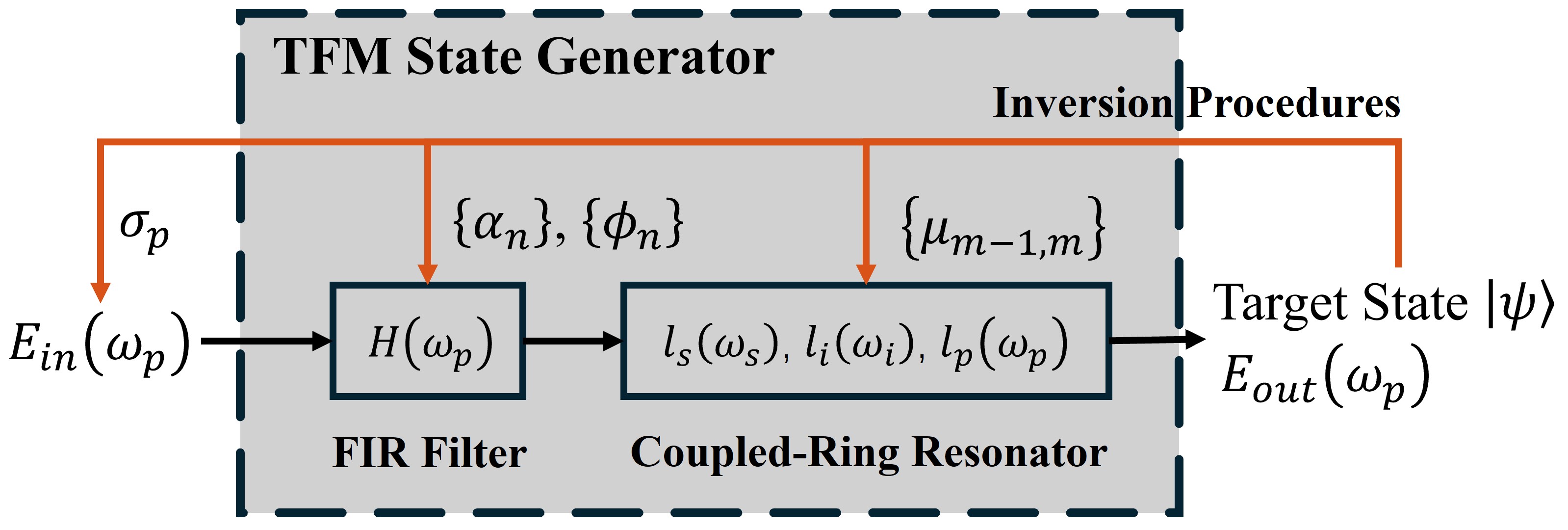}
	\caption{Schematic of the generator's operating principle.}
	\label{fig:3}
\end{figure}
When knowing the spectral distribution of the target TFM-encoded states, inversion procedures are required to determine the physical arrangements of both linear and nonlinear photonic devices. Here, using the state $\left| {{\phi }^{- }} \right\rangle$ as an example, we summarize the steps to determine the model's free parameters:
\begin{enumerate}[itemsep=0pt]
	\item Identify all fixed parameters, including the state's bandwidth ${\sigma}$, the FIR filter's tap number $N$ and base delay time ${\tau}$, the resonant frequencies ${\omega}_{x0}$, and the loss-related constants $1/{{\tau}_{1,x}}$,	$1/{{\tau}_{2,x}}$, ${\kappa}_{x}$.
	\item Specify mutual coupling coefficient ${\mu}_{12}$ for signal and idler resonances and compute $TDSI\left( {{\omega }_{s}},{{\omega }_{i}} \right)$.
	\item Decouple the target JSA of $\left| {{\phi }^{- }} \right\rangle$ from the $TDSI\left( {{\omega }_{s}},{{\omega }_{i}} \right)$ and extract the anti-diagonal distribution of the TDSI-decoupled JSA.
	\item Fit the anti-diagonal distribution to the ADP function in Eq.~(\ref{Eq:5}) using a least-squares algorithm to find all free parameters, including the pump's bandwidth ${\sigma}_p$ and the modulation factors $\left\{ {{\alpha }_{n}} \right\}$ and $\left\{ {{\phi }_{n}} \right\}$ for all $N$ taps. Construct the practical JSA using Eq.~(\ref{Eq:4}) and calculate its state fidelity.
	\item Perform step (4) iteratively for various initial ${\phi }_{n}$ combinations to optimize state fidelity.
	\item Specify another value of ${\mu}_{12}$ and repeat steps (3)-(5).
\end{enumerate}
In steps (2) and (3), the TDSI function's influence on the target state is significantly reduced, allowing the inverse problem to be simplified from two dimensions to a one-dimensional problem, governed by the probability amplitude distribution along the anti-diagonal direction. As a result, steps (4) and (5) can obtain high state fidelity at a low computational cost. For high-dimensional states $\left| {{\Psi }_{D=3}}  \right\rangle$ and $\left| {{\Psi }_{D=4}}  \right\rangle$, the number of stages M in Fig.~\ref{fig:1} (b) is increased to match the dimension of the target state, enabling more precise control over the TDSI function, and additional optimization of coupling coefficients such as ${\mu}_{23}$ and ${\mu}_{34}$ are required. Notably, the same generator can generate lower-dimensional states by detuning the extra rings. To summarize, Figure~\ref{fig:3} sketches the generator's operating principle.
\section{Simulations}
\begin{figure}[t]
	\centering
	\includegraphics[width=\linewidth]{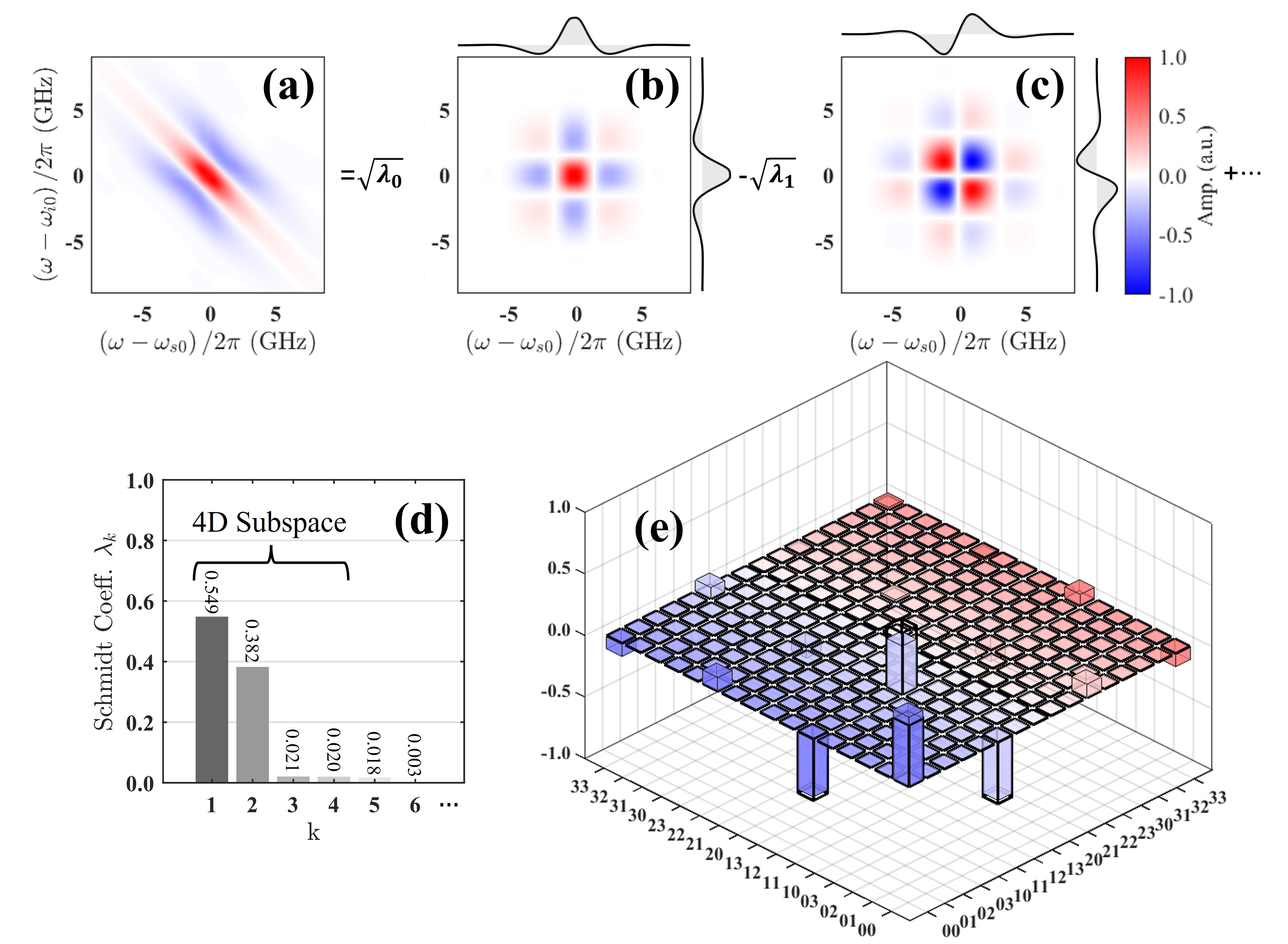}
	\caption{(a) JSA of the generated biphoton state $\left| {{\phi }^{- }} \right\rangle$, (b)-(d) its Schmidt decomposition, and (e) the practical (colored bars) and ideal (clear bars) density matrices.  The ideal density matrix has four bars of equal 0.5 height, and all the other matrix elements are zero.}
	\label{fig:4}
\end{figure}
\begin{figure}[t]
	\centering
	\includegraphics[width=\linewidth]{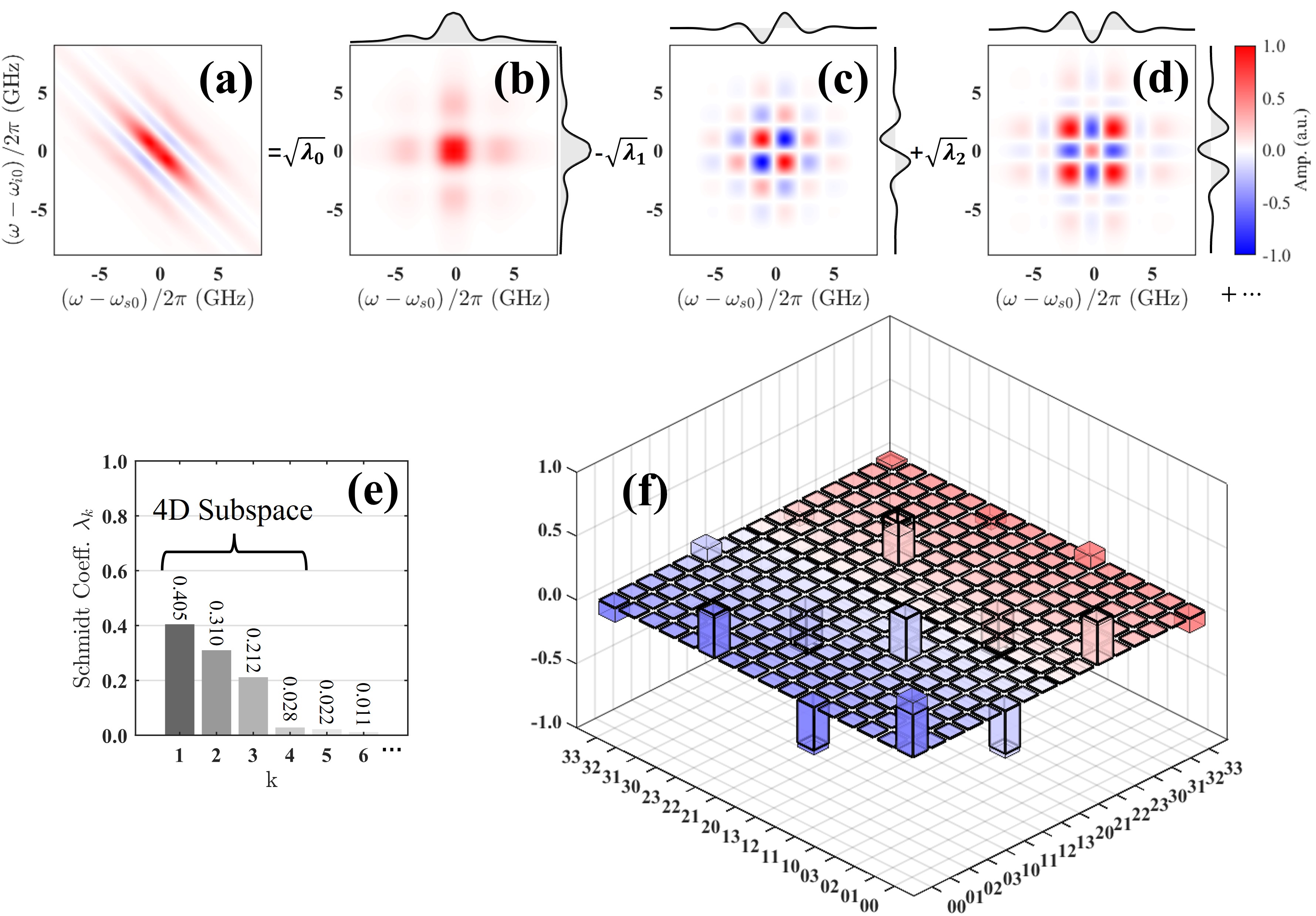}
	\caption{(a) JSA of the generated biphoton state $\left| {{\Psi }_{D=3}}  \right\rangle$, (b)-(e) its Schmidt decomposition, and (f) the practical (colored bars) and ideal (clear bars) density matrices.  The ideal density matrix has nine bars of equal 1/3 height, and all the other matrix elements are zero.}
	\label{fig:5}
\end{figure}

\begin{figure}[t]
	\centering
	\includegraphics[width=\linewidth]{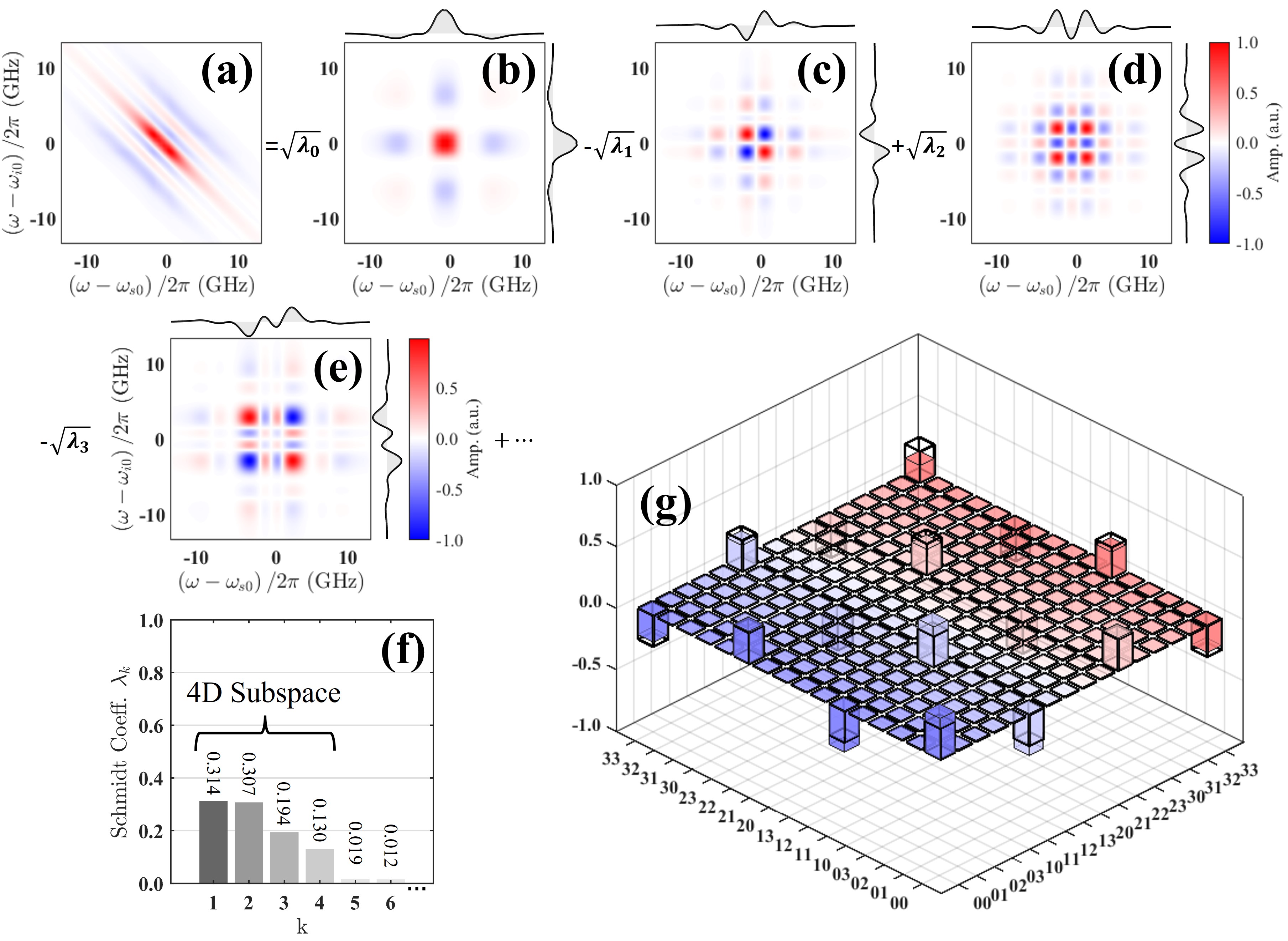}
	\caption{(a) JSA of the generated biphoton state $\left| {{\Psi }_{D=4}}  \right\rangle$, (b)-(f) its Schmidt decomposition, and (g) the practical (colored bars) and ideal (clear bars) density matrices. The ideal density matrix has sixteen bars of equal 1/4 height, and all the other matrix elements are zero.}
	\label{fig:6}
\end{figure}

The fixed parameters for all three target states in Eq.~\ref{Eq:6} are listed in Table~\ref{Tb:S1} in Appendix C. Given that the practical JSA calculated from Eq.~\ref{Eq:4} contains nonlinear phase terms, we impose an $\exp \left( i\pi  \right)$ phase shift at each minimum along the anti-diagonal of $\left| \text{JSA} \right|$. A previous work has confirmed the effectiveness of this phase-imposing method\cite{graffitti2020direct}. Figures~\ref{fig:4}(a), \ref{fig:5}(a), and \ref{fig:6}(a) show the JSAs of the optimized states $\left| {{\phi }^{- }} \right\rangle$, $\left| {{\Psi }_{D=3}}  \right\rangle$, and $\left| {{\Psi }_{D=4}}  \right\rangle$, respectively, with their free parameters listed in Table~\ref{Tb:S2} in Appendix C. The time-frequency entanglement of photon pairs is characterized via Schmidt decomposition of the JSA, as shown in Figs.~\ref{fig:4}(b)-\ref{fig:4}(d), \ref{fig:5}(b)-\ref{fig:5}(e), and \ref{fig:6}(b)-\ref{fig:6}(e) for the states $\left| {{\phi }^{- }} \right\rangle$, $\left| {{\Psi }_{D=3}}  \right\rangle$, and $\left| {{\Psi }_{D=4}}  \right\rangle$, respectively. Due to the inevitable spectral anti-correlation between the emitted photon pairs\cite{paesani2020near}, each TFM pair features small-amplitude oscillatory side modes at the edges of its joint spectrum, while the central regions exhibit significant HG function characteristics. The density matrices $\hat{\rho }=\left| \psi  \right\rangle \left\langle  \psi  \right|$ for the practical (${{\widehat{\rho }}_{pra}}$) and ideal (${{\widehat{\rho }}_{ideal}}$) states, as shown in Fig.~\ref{fig:4}(e) for $\left| {{\phi }^{- }} \right\rangle$, Fig.~\ref{fig:5}(f) for $\left| {{\Psi }_{D=3}}  \right\rangle$, and Fig.~\ref{fig:6}(g) for $\left| {{\Psi }_{D=4}}  \right\rangle$ are confined to the subspace spanned by the first four TFM pairs, with the cumulative weights of higher order pairs being negligible: 0.028 for $\left| {{\phi }^{- }} \right\rangle$, 0.045 for $\left| {{\Psi }_{D=3}}  \right\rangle$, and 0.055 for $\left| {{\Psi }_{D=4}}  \right\rangle$. The state fidelity, defined as ${{F}_{s}}={{[Tr\left( \sqrt{\sqrt{{{\widehat{\rho }}_{pra}}}{{\widehat{\rho }}_{ideal}}\sqrt{{{\widehat{\rho }}_{pra}}}} \right)]}^{2}}$\cite{feng2022transverse}, is 0.950, 0.954, and 0.971 for $\left| {{\phi }^{- }} \right\rangle$, $\left| {{\Psi }_{D=3}}  \right\rangle$, and $\left| {{\Psi }_{D=4}}  \right\rangle$, respectively. The practical Schmidt number $K'=1/\sum{\lambda _{k}^{2}}$\cite{eberly2006schmidt} differs slightly from the target state's Schmidt number ${K}$, with ${{K}'}/{{K}}=2.23/2.00$ for $\left| {{\phi }^{- }} \right\rangle$, $3.26/3.00$ for $\left| {{\Psi }_{D=3}}  \right\rangle$, and $4.03/4.00$ for $\left| {{\Psi }_{D=4}}  \right\rangle$. The non-unity fidelity and Schmidt number deviation result from the degenerate SFWM process's broadband phase-matching function\cite{paesani2020near}, as well as mismatches between practical ADP and TDSI functions and their ideal counterparts.
\section{Discussion and conclusion}
The pair generation rate (PGR) of the resonator-based source and the losses from the pulse shaper are estimated. For PGR estimation, we assume the least favorable conditions, where the field enhancement of all resonances splits, as described in Eq.~\ref{Eq:3}. This results in a lower total quality factor than the unsplit case. Using the PGR expression (Eq.~\ref{Eq:S4} in Appendix D) for pulsed pump input, as described in Ref. \cite{wu2022optimization}, with a 500 MHz repetition rate and a 1 mW average input power at the resonator's input port, the PGR values for the states $\left| {{\phi }^{- }} \right\rangle$,  $\left| {{\Psi }_{D=3}}  \right\rangle$, and $\left| {{\Psi }_{D=4}}  \right\rangle$ are estimated to be 354.3, 157.4, and 88.3 kHz, respectively, which are viable for the TFM-encoded states' measurement\cite{graffitti2020direct}. Notably, because pulse shaping occurs before photon pair generation, losses from the pulse shaper can be compensated for by increasing the average output power of the pulsed laser. Using rib waveguides for delay lines and edge couplers\cite{siew2021review} can reduce shaping loss to less than 15 dB. As a result, an average power of tens of milliwatts at the laser is sufficient to achieve detectable PGR, indicating that the proposed generator is feasible for practical implementations. In terms of programmability, the generator can also be configured to produce nearly uncorrelated photons with a spectral purity of 0.968, as demonstrated in Appendix E.

In conclusion, we have presented a method for the programmable generation of TFM-encoded quantum states of light on the SOI platform without post-manipulation. The proposed generator is capable of producing TFM-encoded maximally entangled states in two, three, and four dimensions, with simulated state fidelities exceeding 0.95. We anticipate that this method will have an impact on the design and implementation of integrated TFM-encoded state generators.

\section*{ACKNOWLEDGMENT}
This work was supported by the National Key Research and Development Program of China (Grants No. 2021YFB2800201), the National Natural Science Foundation of China (Grants No. U22A2082), and the Ningbo Science and Technology Program (Grants No. 2023Z073).

\section*{AUTHOR DECLARATIONS}
\subsection*{Conflict of Interest}
The authors have no conflicts to disclose.

\section*{DATA AVAILABILITY}
The data that support the findings of this study are available from the corresponding authors upon reasonable request.

\setcounter{equation}{0}
\renewcommand{\theequation}{A\arabic{equation}}
\section*{APPENDIX A: Design for Phase-Matching Function}
The phase matching function is given by\cite{garay2007photon}
\begin{eqnarray}
	\label{Eq:S1}
	{{\phi }_{PM}}\left( {{\omega }_{p}},{{\omega }_{s}},{{\omega }_{i}} \right) &= \text{sinc}\left[ \frac{L}{2}\Delta k\left( {{\omega }_{p}},{{\omega }_{s}},{{\omega }_{i}} \right) \right]\nonumber \\
	& \times \exp \left[ i\frac{L}{2}\Delta k\left( {{\omega }_{p}},{{\omega }_{s}},{{\omega }_{i}} \right) \right],
\end{eqnarray}
where ${L}$ is the length of the waveguide. The phase mismatch factor $\Delta k\left( {{\omega }_{p}},{{\omega }_{s}},{{\omega }_{i}} \right)$ is defined as
\begin{eqnarray}
	\label{Eq:S2}
	\Delta k\left( {{\omega }_{p}},{{\omega }_{s}},{{\omega }_{i}} \right) &=k\left( {{\omega }_{p}} \right)+k\left( {{\omega }_{s}}+{{\omega }_{i}}-{{\omega }_{p}} \right)\nonumber\\
	&-k\left( {{\omega }_{s}} \right)-k\left( {{\omega }_{i}} \right)-\gamma_0 P, 
\end{eqnarray}
where $\gamma_0$ represents the nonlinear parameter including both self-phase and cross-phase modulation, and $P$ is the peak power of the incident pulsed pump. $k\left( {{\omega }_{p}}\right)$, $k\left( {{\omega }_{s}}+{{\omega }_{i}}-{{\omega }_{p}} \right)$, $k\left( {{\omega }_{s}}\right)$, and $k\left( {{\omega }_{i}}\right)$ are wavevectors of four interacting fields. The wavevectors are expanded in a first-order Taylor series at perfectly phase-matched frequencies $\omega_{x}^{0}$ (where ${x}=p,s,i$ for the degenerate pump regime). A detailed derivation of the linear approximation for $\Delta k$ can be found in Ref. \cite{garay2007photon}. In this approximation, all zeroth-order terms vanish at the given incident peak power $P_{0}$, while first-order terms remain. Hence, ${\Delta k}$ approximates a function of the form $\Delta {{k}_{lin}}\left( {c_1}{{\omega }_{s}}+{c_2}{{\omega }_{i}} \right)$ in $\left\{ {{\omega }_{s}},{{\omega }_{i}} \right\}$ space, where ${c_1}$ and ${c_2}$ are constants determined by the first-order coefficients in the expansion at $\omega _{x}^{0}$. When ${c_1}=1$ and ${c_2}={-1}$, the phase matching function ${{\phi }_{PM}}\left( {{\omega }_{s}}-{{\omega }_{i}} \right)$ satisfies the symmetric group velocity matching (sGVM) condition, enabling the source to generate high-dimensional TFM-encoded entangled states\cite{ansari2018tailoring}. 
\begin{figure}[h]
	\centering
	\includegraphics[width=\linewidth]{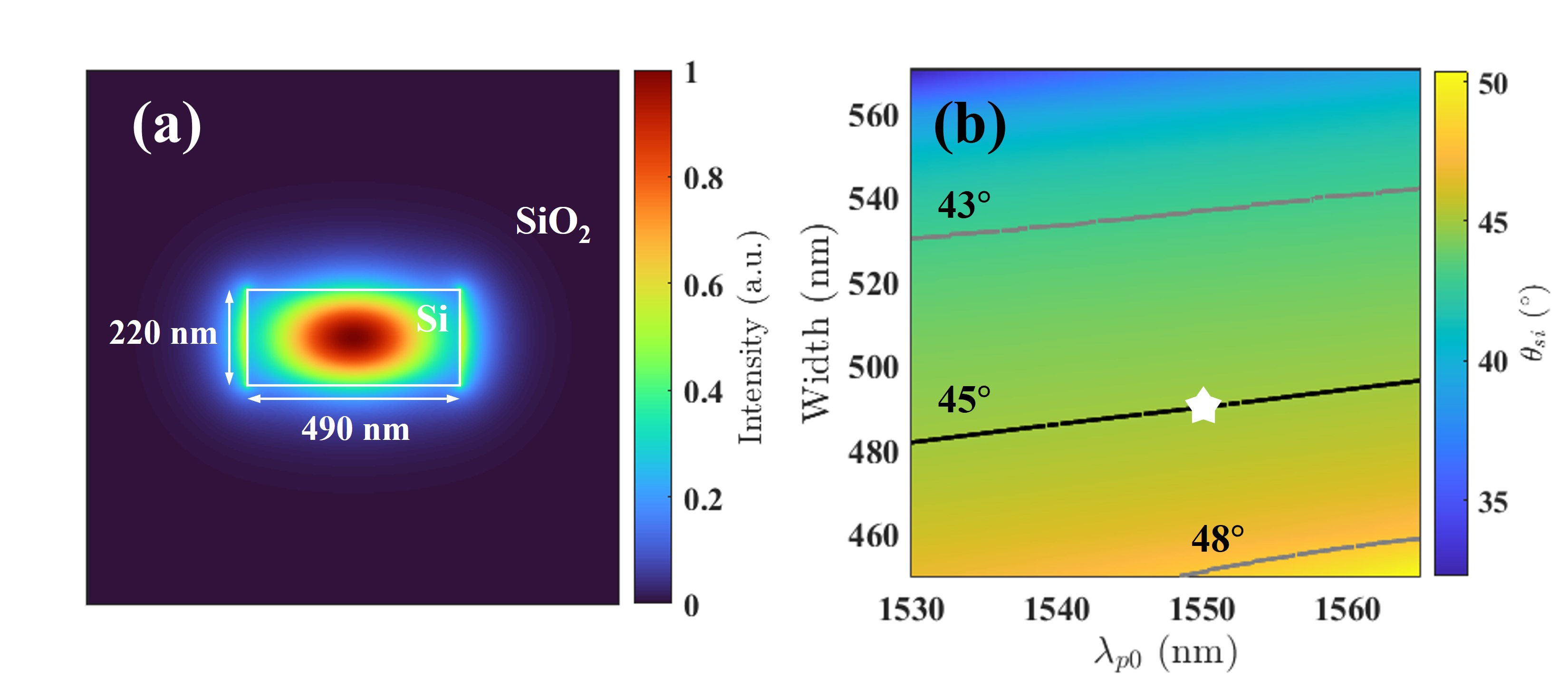}
	\caption{(a) Cross-section of a silicon waveguide showing the ${TE}_{0}$ modal field intensity at 1550 nm. (b) Simulated ${{\theta }_{si}}$ as a function of waveguide width and pump central wavelength.}
	\label{fig:s1}
\end{figure}

To control the time-frequency mode number of the generated state and satisfy the sGVM condition, we engineer the dispersion of an SOI strip waveguide with a commonly used silicon film thickness of 220 nm \cite{siew2021review}. Using the linear approximation method\cite{law2000continuous,garay2007photon,ansari2018tailoring}, the phase-matching orientation angle ${{\theta }_{si}}$, expressed as ${{\theta }_{si}}=-\arctan \left( {{c}_{1}}/{{c}_{2}} \right)$, approximates a fixed value within a few terahertz centered on the pump's carrier frequency, where the group-velocity dispersion is sufficiently small to be neglected. Figure~\ref{fig:s1}(a) shows the cross-section of a typical waveguide, with the fundamental TE modal field intensity simulated using Ansys Lumerical MODE. Numerical simulations show the dependency of ${{\theta }_{si}}$ on both waveguide width and pump central wavelength, as shown in Fig.~\ref{fig:s1}(b). For the photon-pair source, we choose a waveguide width of 490 nm and a pump's central wavelength of about 1550 nm (indicated by the starred data point in Fig.~\ref{fig:s1}(b)). This configuration corresponds to a ${{\theta }_{si}}$ of approximately ${{45}^{\circ}}$, which is optimal for achieving a phase-matching function of the form ${{\phi }_{PM}}\left( {{\omega }_{s}}-{{\omega }_{i}} \right)$.

\setcounter{equation}{0}
\renewcommand{\theequation}{B\arabic{equation}}
\section*{APPENDIX B: Spectral Hermite-Gaussian Functions}
The spectral Hermite-Gaussian functions, providing a complete set of orthogonal bases to decompose the complex electric field of photons, can be expressed as\cite{brecht2015photon}
\begin{equation}
	\label{Eq:S3}
	{{f}_{n}}\left( \Delta {{\omega }_{x}} \right)=\frac{1}{\sqrt{n!\sqrt{2\pi }{{2}^{n}}\sigma }}{{H}_{n}}\left( \frac{\Delta {{\omega }_{x}}}{\sigma } \right)\exp \left[ -\frac{1}{2}{{\left( \frac{\Delta {{\omega }_{x}}}{\sigma } \right)}^{2}} \right],
\end{equation}
where $\Delta {{\omega }_{x}}={{\omega }_{x}}-{{\omega }_{x0}}$ is the detuning between a spectral component ${{\omega }_{x}}$ of the $x$ photon and its neighboring resonant frequency ${{\omega }_{x0}}$, $2{\sigma}$ is the full width at $1/e$ maximum of the probability distribution of the photon, and ${{H}_{n}}\left( \Delta {{\omega }_{x}}/\sigma  \right)$ is the n-th order Hermite polynomial in the variable $\Delta {{\omega }_{x}}/\sigma$. Notably, the functions ${{f}_{n}}\left( {x} \right)$ adhere to the normalization condition ${{\int_{-\infty }^{\infty }{\left| {{f}_{n}}\left( x \right) \right|}}^{2}}dx=1$, and their positive and negative values correspond to a $\pi$ phase shift.

\section*{APPENDIX C: Parameters}
All parameters are listed in Table~\ref{Tb:S1} and~\ref{Tb:S2}.
\begin{table*}[hp]
	\centering
	\caption{\label{Tb:S1}\bf Model's fixed parameters.}
	\begin{tabular}{
			@{\extracolsep{0.05cm}} 
			c c c c c c c
		}
		\hline \hline
		Parameter & N & $\tau$ (ps) & ${\omega}_{i0}$ (THz) & $1/{{\tau}_{1,i}}$ (GHz) & $1/{{\tau}_{m,i}}$ (GHz) & ${\omega}_{p0}$ (THz)\\
		Value ($\left| {{\phi }^{- }} \right\rangle$) & 6 & 75 & 1214.45 & 7.26 & 2.44 & 1215.07\\
		Value ($\left| {\Psi}_{D=3}  \right\rangle$) & 6 & 75 & 1214.45 & 9.68 & 2.44 & 1215.07\\
		Value ($\left| {\Psi}_{D=4}  \right\rangle$) & 6 & 75 & 1214.45 & 12.11 & 2.44 & 1215.07\\
		Value (separable state) & 6 & 75 & 1214.45 & 7.26 & 2.44 & 1215.07\\
		\hline
		Parameter & $1/{{\tau}_{1,p}}$ (GHz) & $1/{{\tau}_{m,p}}$ (GHz) & ${\omega}_{s0}$ (THz) & $1/{{\tau}_{1,s}}$ (GHz) & $1/{{\tau}_{m,s}}$ (GHz) & ${\kappa}$ ($\sqrt{\text{THz}}$)\\
		Value ($\left| {{\phi }^{- }} \right\rangle$) & 7.26 & 2.44 & 1215.70 & 7.26 & 2.44 & 0.0985\\
		Value ($\left| {\Psi}_{D=3}  \right\rangle$) & 9.68 & 2.44 & 1215.70 & 9.68 & 2.44 & 0.1206\\
		Value ($\left| {\Psi}_{D=4}  \right\rangle$) & 12.11 & 2.44 & 1215.70 & 12.11 & 2.44 & 0.1393\\
		Value (separable state) & 7.26 & 2.44 & 1215.70 & 7.26 & 2.44 & 0.0985\\
		\hline\hline
	\end{tabular}
\end{table*}
\begin{table*}[hp]
	\centering
	\caption{\label{Tb:S2}\bf Model's free parameters.}
	\begin{tabular}{
			@{\extracolsep{0.05cm}} 
			c c c c c c c c c
		}
		\hline \hline
		Parameter & ${\sigma}_{p}$ (GHz) & ${\mu}_{12,i}$ (GHz) & ${\mu}_{12,p}$ (GHz) & ${\mu}_{12,s}$ (GHz) & ${\mu}_{23,i}$ (GHz)\\
		Value ($\left| {{\phi }^{- }} \right\rangle$) & $4.02*{2\pi}$ & 1.45 & 0 & 1.45 & /\\
		Value ($\left| {\Psi}_{D=3}  \right\rangle$) & $2.14*{2\pi}$ & 2.76 & 0 & 2.76 & 1.66\\
		Value ($\left| {\Psi}_{D=4}  \right\rangle$) & $2.96*{2\pi}$ & 3.88 & 0 & 3.88 & 3.10\\
		Value (separable state) & $60*{2\pi}$ & 0 & 6.30 & 0 & /\\
		\hline
		Parameter & ${\mu}_{23,p}$ (GHz) & ${\mu}_{23,s}$ (GHz) & ${\mu}_{34,i}$ (GHz) & ${\mu}_{34,p}$ (GHz) & ${\mu}_{34,s}$ (GHz)\\
		Value ($\left| {{\phi }^{- }} \right\rangle$) & / & / & / & / & /\\
		Value ($\left| {\Psi}_{D=3}  \right\rangle$) & 0 & 1.66 & / & / & /\\
		Value ($\left| {\Psi}_{D=4}  \right\rangle$) & 0 & 3.10 & 2.48 & 0 & 2.48\\
		Value (separable state) & / & / & / & / & /\\
		\hline
		Parameter & ${\alpha}_{1}$ & ${\phi}_{1}$ & ${\alpha}_{2}$ & ${\phi}_{2}$ & ${\alpha}_{3}$ & ${\phi}_{3}$\\
		Value ($\left| {{\phi }^{- }} \right\rangle$) & 0.216 & 1.590 & 0.805 & 1.187 & 0.123 & 2.670\\
		Value ($\left| {\Psi}_{D=3}  \right\rangle$) & 0.858 & 6.282 & 0.701 & 4.302 & 0 & 0.076\\
		Value ($\left| {\Psi}_{D=4}  \right\rangle$) & 0.840 & 3.568 & 0.970 & 2.339 & 0.580 & 5.971\\
		Value (separable state) & 1 & 0 & 0 & 0 & 0 & 0\\
		\hline
		Parameter & ${\alpha}_{4}$ & ${\phi}_{4}$ & ${\alpha}_{5}$ & ${\phi}_{5}$ & ${\alpha}_{6}$ & ${\phi}_{6}$ \\
		Value ($\left| {{\phi }^{- }} \right\rangle$) & 0.995 & 4.860 & 0.754 & 1.108 & 0.523 & 6.282\\
		Value ($\left| {\Psi}_{D=3}  \right\rangle$) & 0.983 & 3.210 & 0.358 & 1.013 & 0.579 & 2.281\\
		Value ($\left| {\Psi}_{D=4}  \right\rangle$) & 0.768 & 4.800 & 0.741 & 5.451 & 0.910 & 4.951\\
		Value (separable state) & 0 & 0 & 0 & 0 & 0 & 0\\
		\hline\hline
	\end{tabular}
\end{table*}

\setcounter{equation}{0}
\renewcommand{\theequation}{C\arabic{equation}}
\section*{APPENDIX D: Estimation for Pair Generation Rate}
The shaped pulses and the coupled-ring resonator's pump resonance have comparable bandwidths, so the generated pairs per pulse can be estimated as \cite{wu2022optimization}:
\begin{equation}
	\label{Eq:S4}
	{{N}_{one-pulse}}=\frac{3{{\gamma }^{2}}{{W}^{2}}V_{g}^{4}}{8{{\pi }^{2}}{{R}^{2}}\omega _{p0}^{2}}\frac{Q_{tot}^{6}}{Q_{ext}^{4}},
\end{equation}
where $V_{g}=7.14\times {{10}^{7}} m/s$ is the group velocity and $R$ is the radius of the main ring. The nonlinear parameter $\gamma$ is defined as $\gamma ={{n}_{2}}{{\omega }_{p0}}/\left( c{{A}_{eff}} \right)$, where ${{n}_{2}}=5.59\times {{10}^{-18}}{{m}^{2}}/W$ represents the nonlinear Kerr index of silicon at 1550 nm \cite{zhang2014nonlinear}, $c$ is the speed of light in a vacuum, and ${{A}_{eff}}=0.191\times {{10}^{-12}}{{m}^{2}}$ represents the effective mode area of the waveguide. The resonator's total and extrinsic quality factors are defined as ${{Q}_{tot}}={{\omega }_{p0}}/\left( 2\sum\limits_{m=1}^{M}{1/{{\tau }_{m,p}}} \right)$ and ${{Q}_{ext}}={{\omega }_{p0}}/{{\kappa }^{2}}$, respectively. The single-pulse energy $W$ is defined as $W={P_{avg}}/{R_{rep}}$, where ${P_{avg}}=1~ \text{mW}$ is the average power of the pulsed pump at the resonator's input port, and ${R_{rep}}=500~\text{MHz}$ is the repetition rate. Using Eq.~\ref{Eq:S4}, the PGR values for the states $\left| {{\phi }^{- }} \right\rangle$,  $\left| {{\Psi }_{D=3}}  \right\rangle$, and $\left| {{\Psi }_{D=4}}  \right\rangle$ are estimated to be 354.3, 157.4, and 88.3 kHz, respectively.

\section*{APPENDIX E: Separable State Generation}
\begin{figure}[h]
	\centering
	\includegraphics[width=\linewidth]{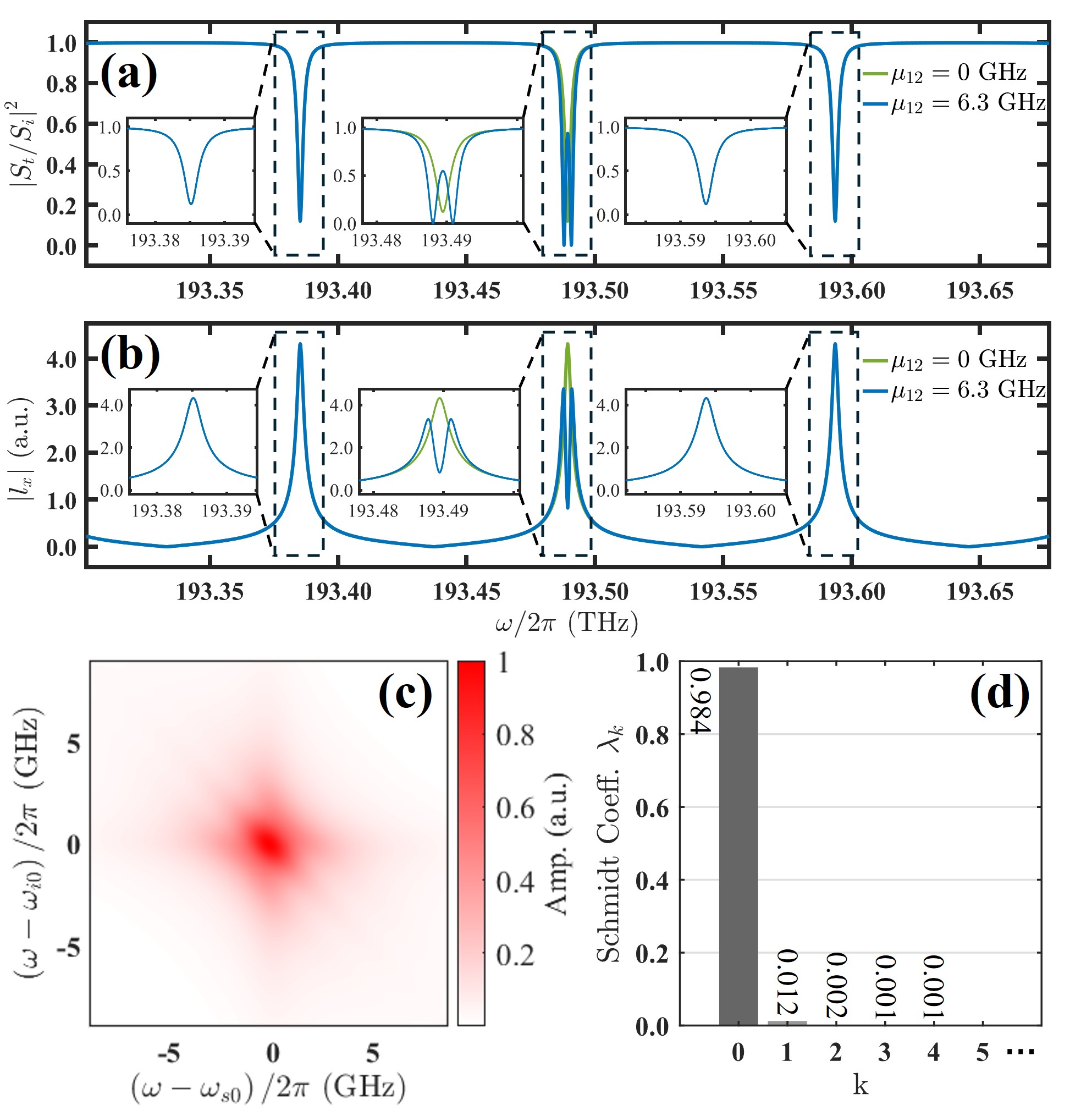}
	\caption{(a) and (b) Spectra of (a) ${{\left| {{S}_{t}}/{{S}_{i}} \right|}^{2}}$ and (b) ${{\left| {l}_{x} \right|}}$ at ${\mu}_{12}$ values of 0 and 6.30 GHz. Insets provide magnified views of the spectra at the resonant frequencies. (c) JSA of the generated biphoton separable state, and (d) the weights of the first six Schmidt modes.}
	\label{fig:s2}
\end{figure}

Adding an extra $\pi$ phase shift to the phase shifter ${{\phi }_{h3}}$ of the MZI coupler shown in Fig. 2(a) converts the signal (idler)-split scenario shown in Figs. 2(c) and 2(d) into a distinct pump-split scenario shown in Figs.~\ref{fig:s2}(a) and~\ref{fig:s2}(b). The generation of a nearly separable state follows from this pump-split scenario. High spectral purity is achieved by using only the first tap of the FIR filter and increasing the bandwidth $2{\sigma}_{p}$. The remaining step in optimizing purity involves determining the optimal coupling coefficient ${\mu}_{12}$, which is optimized at 6.30 GHz. The values of ${\mu}_{12}$ and other free parameters are provided in Table~\ref{Tb:S2} . The JSA for the generated state is shown in Fig.~\ref{fig:s2}(c), with a spectral purity $P=\sum\nolimits_{k}{\lambda _{k}^{2}}$ of 0.968, calculated from the weights ${\lambda _{k}}$ of the first six Schmidt modes, as shown in Fig.~\ref{fig:s2}(d). 

When compared to high-spectral-purity single-photon sources based on dual-pulse configurations \cite{Christensen:18,burridge2020high} or interferometrically coupled resonators \cite{vernon2017truly,liu2019high,burridge2023integrate,alexander2024manufacturable}, the proposed state generator does not offer a higher spectral purity upper limit. The theoretical upper limit of 0.97, demonstrated in Ref. \cite{mccutcheon2021backscattering}, is consistent with our simulation results and exceeds the typical purity of 0.93\cite{vernon2017truly} achievable with a conventional ring resonator that does not exhibit pump resonance splitting. However, our scheme maintains a notable advantage in terms of reconfigurability as it can generate both TFM-encoded entangled and nearly separable states.

	\bibliographystyle{unsrt}
 	\bibliography{document.bib}
 	
	\end{sloppypar}
\end{document}